\documentclass[aps,pra,preprint,longbibliography,superscriptaddress,endfloats*,11pt]{revtex4-1}

\usepackage{color}
\usepackage{xcolor}
\usepackage{graphicx}
\usepackage{amsmath}
\usepackage{amsfonts}
\usepackage{soul}
\usepackage[normalem]{ulem}
\usepackage{lipsum}
\usepackage{bibentry}
\nobibliography*

\newcommand{\jdedit}[1] {{\textcolor{black}{#1}}}

\begin{document}

\title{Rogue waves and analogies in optics and oceanography}

\author{John M. Dudley}
\affiliation{Institut FEMTO-ST, Universit\'{e} Bourgogne Franche-Comt\'{e} CNRS UMR 6174, 25000 Besan\c{c}on, France}
\author{Go\"{e}ry Genty}
\affiliation{Tampere University, Photonics Laboratory, Fi-33104 Tampere, Finland}
\author{Arnaud Mussot}
\affiliation{Univ. Lille, CNRS, UMR 8523 - PhLAM - Physique des Lasers Atomes et Mol\'{e}cules, F-59000 Lille, France}
\author{Amin Chabchoub}
\affiliation{Centre for Wind, Waves and Water, School of Civil Engineering, The University of Sydney, Sydney, NSW 2006, Australia}
\author{Fr\'{e}d\'{e}ric Dias}
\affiliation{School of Mathematics and Statistics, University College Dublin, MaREI Centre,
Belfield, Dublin 4, Ireland \vspace{3ex}}

\begin{abstract}

\noindent We review the study of rogue waves and related instabilities in optical and oceanic environments, with particular focus on recent experimental developments.  In optics, we emphasize results arising from the use of real-time measurement techniques, whilst in oceanography we consider insights obtained from analysis of real-world ocean wave data and controlled experiments in wave tanks.  Although significant progress in understanding rogue waves has been made based on an analogy between wave dynamics in optics and hydrodynamics, these comparisons have predominantly focused on one-dimensional nonlinear propagation scenarios. As a result, there remains significant debate about the dominant physical mechanisms driving the generation of ocean rogue waves in the complex environment of the open sea. Here, we review state-of-the-art of rogue wave studies in optics and hydrodynamics, aiming to clearly identify similarities and differences between the results obtained in the two fields.  In hydrodynamics, we take care to review results that support both nonlinear and linear interpretations of ocean rogue wave formation, and in optics, we also summarise results from an emerging area of research applying the measurement techniques developed for the study of rogue waves to dissipative soliton systems.  We conclude with a discussion of important future research directions.

\vskip 15mm

\noindent Note: this repository version is the initially submitted manuscript associated with the paper:\\
\noindent Dudley, J.M., Genty, G., Mussot, A. Chabchoub, A, Dias, F.\\
Rogue waves and analogies in optics and oceanography.\\
Nature Reviews Physics 1, 675–689 (2019)\\
doi:10.1038/s42254-019-0100-0

\end{abstract}

\maketitle

\section{Introduction}

Many complex systems in nature show the unexpected emergence of rare and extreme events that can have dramatic impact on their surrounding environment. A well-known example comes from oceanography, where large amplitude ``rogue waves'' have been associated with many maritime disasters, and where the difficulty in understanding their physical origin has made them as much a part of folklore as of science \cite{Kharif-2003,Kharif-2008,Olagnon-2017}.  Research into rogue waves has developed significantly since 2007 following the suggestion of a powerful analogy between the generation of large ocean waves and the propagation of light fields in optical fibre \cite{Solli-2007}.  This analogy was made based on the observation of long-tailed statistics in experiments studying noise in fibre supercontinuum generation, and it attracted intense interest in both optics and hydrodynamics.  Many subsequent studies explored the analogous characteristics of the two systems in more detail, with a particular aim to clarify the potential role of nonlinear focusing effects during propagation \cite{Akhmediev-2009,Dudley-2008,Akhmediev-2011,Akhmediev-2016}.  The study of rogue waves has now been established as a very active field of research \cite{Akhmediev-2013,Onorato-2013,Adcock-2014}.

In optics, however, some  misunderstanding has developed because the description ``optical rogue wave'' has been applied to instabilities in a range of optical systems exhibiting long-tailed statistics, irrespective of any potential hydrodynamic analogy.  In addition, the terminology ``rogue wave'' has become common in a purely mathematical sense to describe strongly localized solutions of certain nonlinear partial differential equations.  In parallel, although nonlinear focussing has received the majority of attention in hydrodynamic studies, there remains significant debate concerning the relative importance of linear and nonlinear effects in actually driving ocean rogue wave generation. There is also related discussion concerning the limitations of one-dimensional models when applied to the complex oceanic environment.  These various issues can make it difficult for non-specialists to appreciate and understand a subject of evident interest.

Our aim here is to provide an overview of the study of rogue waves in optics and hydrodynamics in a way which aims to be accessible to workers from different fields.  Because the analogy between rogue waves in optics and hydrodynamics was initially made based on the suggested importance of nonlinear focussing in both domains, we begin by introducing the main regime in which optical and hydrodynamic nonlinear wave propagation can be described by a common formalism.  We then review a selection of results obtained in different optical and hydrodynamic systems, including also results that have considered purely linear origins for rogue waves. Where possible, we focus on recent experimental results.  In the particular case of optics, after first reviewing systems where a potential analogy exists with ocean wave propagation, we then summarize results in an area where no such analogy exists -- an emerging field where instrumentation developed to study optical rogue waves is being applied to characterize transient dynamics and novel soliton behaviour (dissipative rogue waves) in lasers.  When considering water waves, we review studies of real-world ocean wave data, as well as wave tank experiments in controlled environments. We conclude with a discussion of future research directions.

\section{Nonlinear Focussing in Optics and Hydrodynamics}

The wave equations of optics and hydrodynamics are simplified models derived from Maxwell's equations and the Navier-Stokes equations respectively \cite{Agrawal-2012,Mei-2005,Ablowitz-2011}.  For uni-directional propagation and assuming that underlying carrier waves are modulated by a slowly-varying narrowband envelope, it is possible to derive a common mathematical formalism in the form of a nonlinear Schr\"{o}dinger equation (NLSE) describing the envelope evolution in space and time.  In optics, the NLSE applies to an envelope modulating electromagnetic carrier waves in optical fibre, whilst in hydrodynamics, it applies to an envelope that modulates long-crested surface gravity waves on deep water (and even on water of intermediate depth \cite{Mei-2005}).

The NLSE models the effect of group velocity dispersion and nonlinearity and, since we are interested in mechanisms that can potentially lead to large envelope amplitudes, we consider the regime where dispersion and nonlinearity act together to support ``nonlinear focussing''.  A physical understanding of why optical and water waves are described by the same model can be obtained by associating the origin of the nonlinearity in both cases with a nonlinear dispersion relation. In this picture, optical waves with higher amplitude propagate at reduced velocity whereas water waves with higher amplitude propagate at increased velocity. Since the corresponding linear dispersive terms in optics (with anomalous dispersion) and hydrodynamics have opposite signs, the NLSE describes a focussing mechanism in both cases.

Although the NLSE is a simplified model, it has proven \jdedit{highly} successful in describing wave evolution in specific regimes in optics and hydrodynamics.  Of course, generalized NLSE models have been developed to include higher-order processes \cite{Dysthe-1979,Blow-1989,Dudley-2006,Chabchoub-2013}, but the simple NLSE has remained attractive for analytic and numerical study because it clearly isolates the dispersive and nonlinear contributions to the dynamics. Moreover, the integrability of the NLSE yields a number of analytic solutions that can be used to determine experimental initial conditions to \jdedit{coherently seed} the generation of localized structures considered as prototype rogue waves \cite{Akhmediev-1997}.  Table I gives the dimensional form of the focussing NLSE for optics and hydrodynamics, a commonly encountered normalised form with dimensional transformations, and a convenient listing of typical parameters associated with rogue waves and their measurements in optics and hydrodynamics.

\begin{table}
\begin{center}
\includegraphics[angle=0,origin = c, scale=0.65]{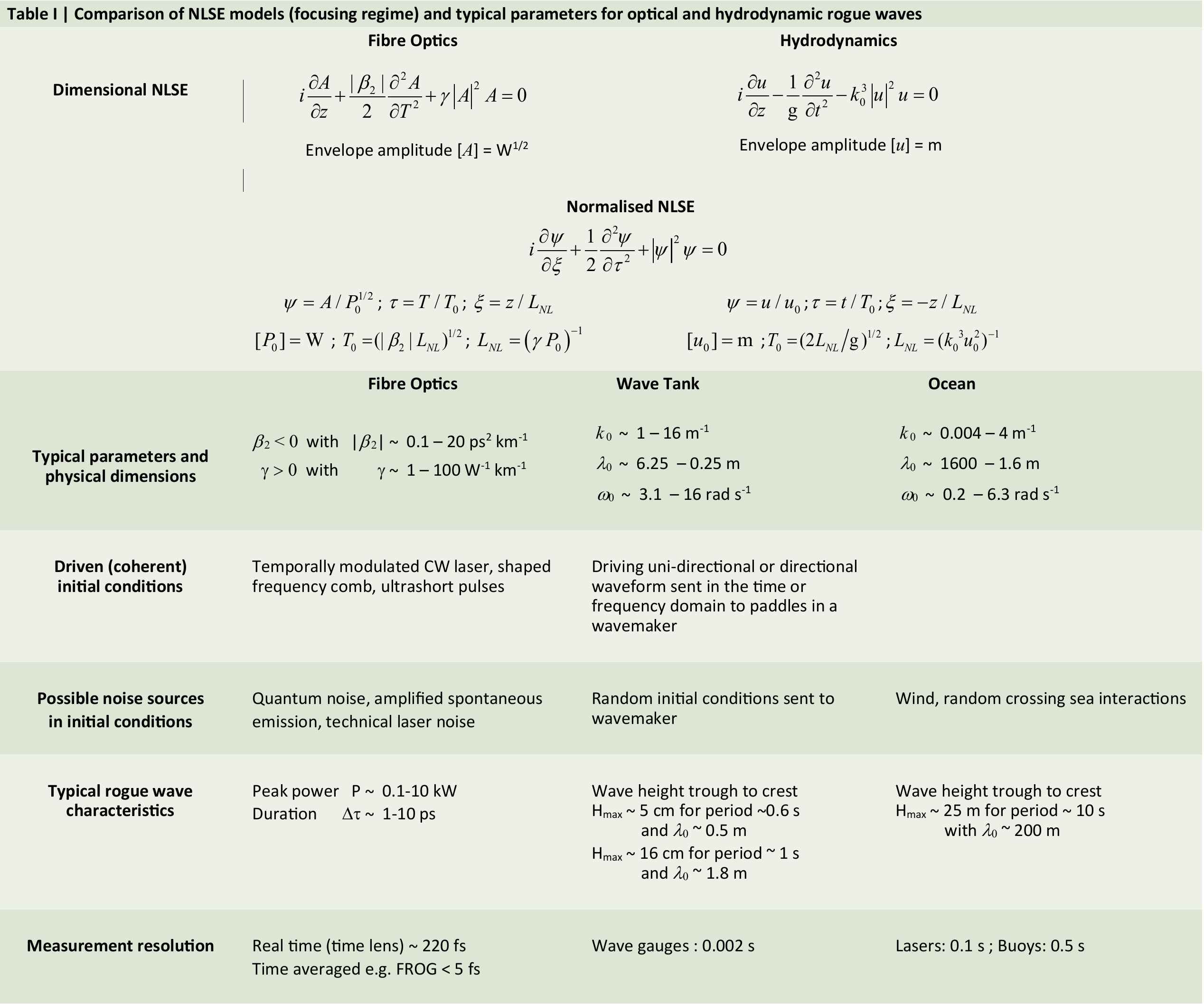}
\caption{Explicit form of normalized and dimensional focusing NLSE models in optics and for surface waves on deep water.  In optics, the NLSE describes the evolution of an optical envelope amplitude $A(z,T)$ while in hydrodynamics it describes the evolution of the envelope $u(z,t)$ of a group of deep water waves.     The table also lists typical parameters associated with optical and hydrodynamic rogue waves.  For optics, the parameters $\beta_2$ and $\gamma$ are the group velocity dispersion and nonlinearity parameter respectively, and for the focussing NLSE, $\beta_2<0$. In hydrodynamics, $g = 9.81 \, \mathrm{m}\,\mathrm{s}^{-2}$ is the acceleration due to gravity, $k_0 = 2 \pi/\lambda$ is the wavevector and $\omega_0$ is the corresponding angular frequency.}
\label{Table1}
\end{center}
\end{table}

The nonlinear focussing mechanism that has received most attention in the study of rogue waves is that of modulation (or Benjamin-Feir) instability, the exponential growth of a small modulation (or noise) on a continuous wave input to the NLSE \cite{Lighthill-1965,Whitham-1965,Bespalov-1966,Benjamin-1967a,Zakharov-1968,Peregrine-1983}.  Plotting in terms of the normalized NLSE (see Table I), Figs 1a,b show scenarios of modulation instability for two different cases corresponding to narrowband input spectra.  Figure 1a plots the evolution of an ``Akhmediev breather,'' a structure periodic in time $\tau$ that develops from an initial coherent sinusoidal modulation on a continuous wave \cite{Akhmediev-1986}. Because of non-ideal initial conditions, the evolution is also periodic with propagation distance $\xi$, a manifestation of the celebrated Fermi-Pasta-Ulam-Tsingou recurrence phenomenon \cite{Fermi-1955,Dauxois-2008}.   In contrast to this regular behaviour, Fig. 1b shows the random evolution observed when the initial continuous wave is perturbed by low amplitude noise.  Here we see approximate $\tau$-periodicity at the reciprocal of the frequency of maximum amplification for the instability \cite{Akhmediev-1997} and random breathing along $\xi$.



In contrast, Fig. 1c displays the qualitatively different evolution observed when the initial conditions consist of an incoherent field with near 100\% amplitude noise, rather than a weakly perturbed continuous wave.  We still see the emergence of strongly localized peaks, but with more erratic trajectories compared to Fig. 1b, and the evolution here is better described in terms of turbulence and the propagation of higher-order background-free solitons \cite{Suret-2016,SotoCrespo-2016}. As we shall describe below, all three cases in Fig.~1 have been studied in experiments.

\begin{figure}
\begin{center}
\includegraphics[angle=0,origin = c, scale=0.8]{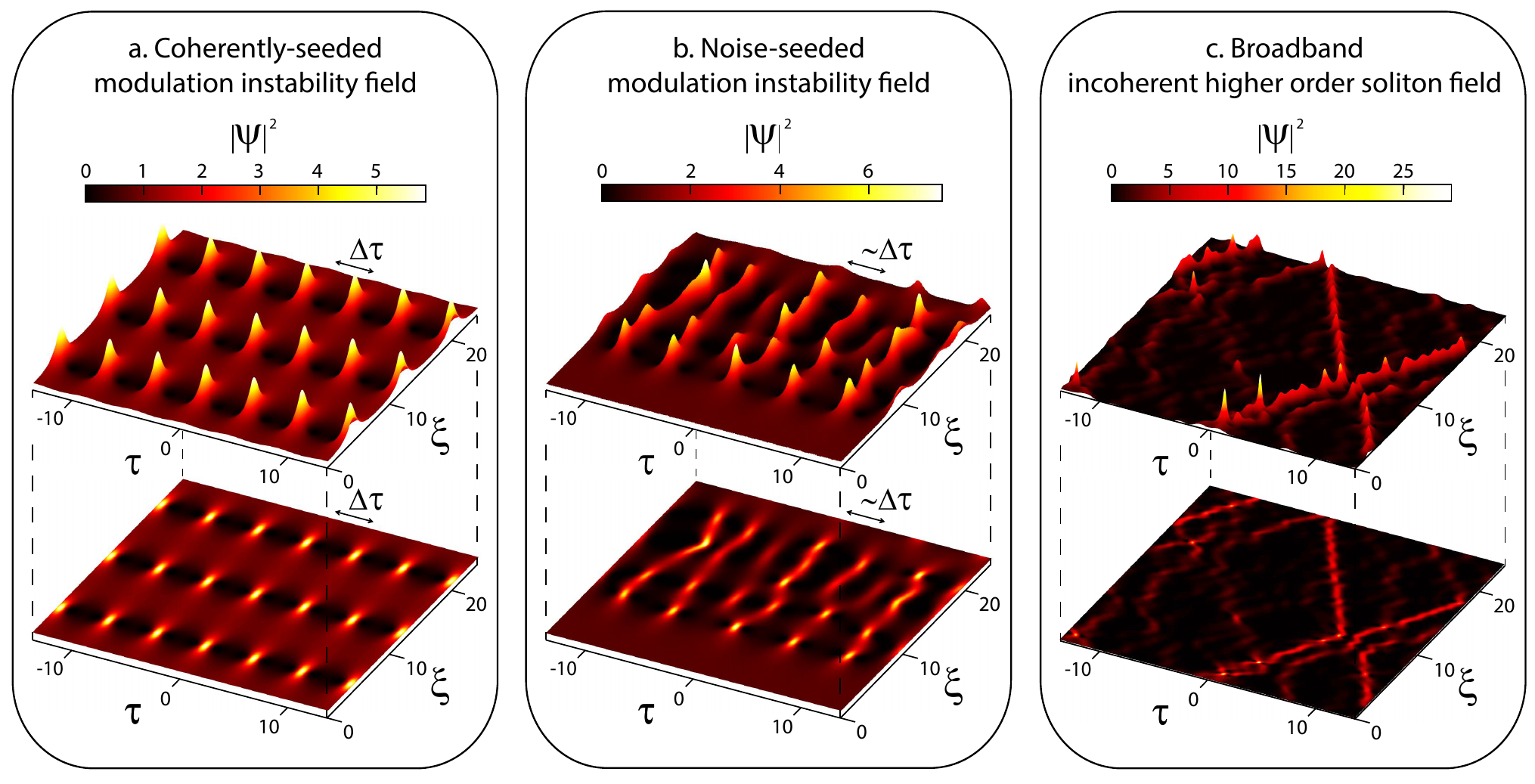}
\caption{Localization properties of nonlinear focussing dynamics in the NLSE for three different cases. \textbf{a} shows ``breathing'' with propagation distance $\xi$ of a train of localized pulses periodic in time $\tau$. Such a train of Akhmediev breathers is generated from an initial coherent modulation on a continuous wave at $\xi = 0$, with modulation frequency at the peak of the instability gain such that the period is $\Delta \tau = \sqrt{2}\pi$.  \textbf{b} shows the complex noisy field of peaks that emerge when the continuous wave initial condition at $\xi = 0$ is perturbed by low amplitude broadband random noise.  In this case we see approximate periodicity with mean period $ \Delta \tau \sim \sqrt{2}\pi$, and random breathing along $\xi$. \textbf{c} shows the qualitatively different behaviour when the initial conditions consist of an incoherent broadband pulse rather than a continuous wave.  Here we see turbulent evolution and the emergence of random higher-order background free solitons.  All plots show the space-time evolution of the intensity $|\psi|^2$ plotted above the corresponding two-dimensional projection.}
\end{center}
\end{figure}

We also remark that, although the form of the NLSE in optics and hydrodynamics is the same, there are significant differences concerning what the model describes in each system. In optics, the underlying carrier wave is considered sinusoidal at frequency $\omega$ whereas in hydrodynamics, the NLSE envelope modulates the Stokes wave which (to second-order) contains contributions at both $\omega$ and the second harmonic $2 \omega$ \cite{Chabchoub-2015}.  In addition, even though the terminology ``optical wave breaking'' is used in an NLSE context to describe the steepening of an optical envelope from nonlinear focussing \cite{Tomlinson-1985}, optical carrier waves themselves do not ``break'' or plunge as they do in hydrodynamics \cite{Babanin-2011,Barthelemy-2018}.   There are also important differences concerning measurements. Specifically, experiments in optical fibres generally measure only the time-domain envelope intensity, and information about carrier oscillations is not recorded.  In contrast, measurements in hydrodynamics directly record the individual carrier wave amplitudes (although envelope information can be reconstructed  straightforwardly \cite{Osborne-2010}). Particular care must therefore be taken when comparing statistics between optics and hydrodynamics, because statistics in fibre optics experiments are determined from the intensity envelope peaks, whereas statistics of water waves are usually determined from the trough-crest heights (or crest heights or amplitudes) of individual waves.  This is of special relevance to the criterion which, in oceanography, identifies a rogue wave as one whose trough-to-crest height exceeds twice the ``significant wave height'' (the mean height of the highest third of waves in a measured population). In optics, although the criterion is the same, it is expressed in terms of envelope peak intensities.

Finally, whilst renewed interest in the link between optics and hydrodynamics was stimulated by the possibility to gain improved understanding of rogue waves, analogous nonlinear propagation effects had in fact been independently observed in both fields over many years.  This is of course to be expected given the centrality of the NLSE in describing nonlinear focussing in both environments, and it is interesting to illustrate in Fig.~2 this parallel development through a timeline both before and after the 2007 optical rogue wave experiments.  Some of these results we discuss in more detail below, and more exhaustive  historical treatments appear in Refs. \cite{Kharif-2008,Zakharov-2009, Agrawal-2012, Onorato-2013,Adcock-2014,Dudley-2014}.

\begin{figure}
\begin{center}
\includegraphics[angle=0,origin = c, scale=0.65]{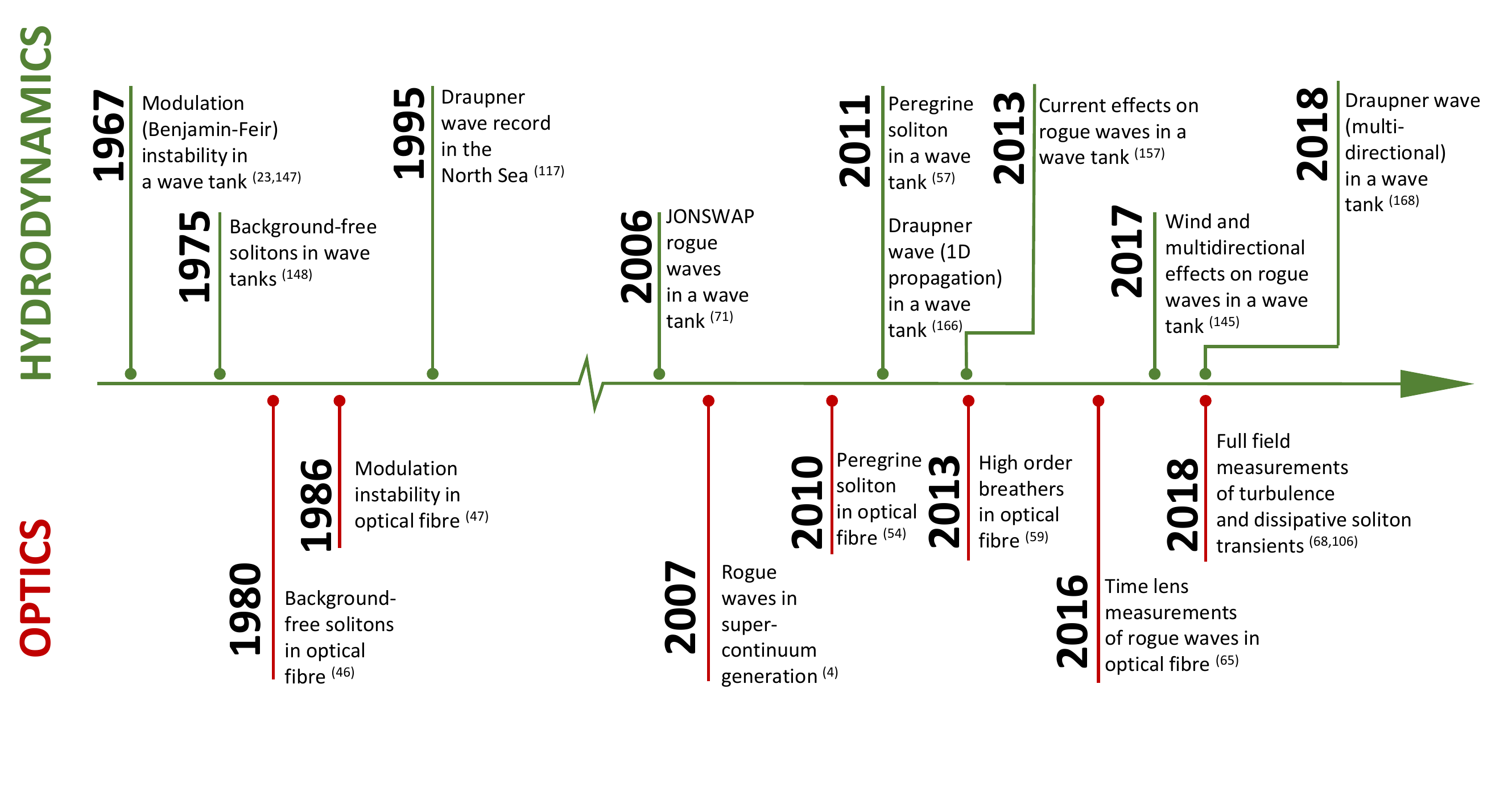}
\caption{Timeline illustrating the parallel development of major experimental findings in hydrodynamics (top) and fibre optics (bottom).  The references given in the figure refer to the bibliography. }
\end{center}
\end{figure}

\section{Rogue waves in optics}

\subsection{Overview}

As mentioned above, the 2007 proposal that an optical system could display properties mimicking oceanic rogue waves was  made based on noise measurements of an optical supercontinuum \cite{Solli-2007}. Specifically, a relatively new experimental technique for the time known as the dispersive Fourier transform (DFT) was used to measure pulse-to-pulse (``shot-to-shot'') fluctuations in the shape of supercontinuum spectra generated from high-power picosecond pulses injected in an optical fibre.  Although it was well-known that stable supercontinuum generation was possible using femtosecond pulse pumping, the DFT measurements of a picosecond supercontinuum revealed dramatic variations in the structure of the spectra generated from sequential pulses.  Moreover, when filtering the supercontinuum spectra at long wavelengths where distinct (background free) optical solitons were expected to form, the corresponding time series showed a highly asymmetric ``L-shaped'' probability distribution, with a long tail containing a small population of high intensity peaks.

A link between these spectral instabilities and ocean rogue waves was proposed based firstly on the similarity of these long-tailed statistics to those of extreme events \cite{Kotz-2000}, and also because of the common NLSE model for deep water waves and optical fibre propagation discussed above.  The supercontinuum results attracted immediate interest and stimulated many other experiments in optics.  However, the description ``optical rogue wave'' was rapidly generalized to refer to any optical system displaying long tailed statistics, even when there was no obvious correspondence with analogous oceanic dynamics.

Although this broader usage in optics is now well-established, to avoid confusion it is essential to specify when a particular experiment in optics has a potential analogy with a hydrodynamic or oceanographic system. In what follows, we focus firstly on describing one and two dimensional propagation experiments where there is indeed a potential hydrodynamic analogy, and we then review the rapidly emerging field of real-time characterisation of dissipative soliton instabilities in lasers.  Although these laser dynamics are likely without a natural oceanographic equivalent, it is a field of study that has developed directly from applying the real-time measurement techniques used for the study of rogue waves in optics, and it may well stimulate studies of other classes of nonlinear dissipative structures in hydrodynamics.

\subsection{Nonlinear Rogue Waves in Optical Fibre Systems}

Following the initial experiments by Solli \emph{et al.} \cite{Solli-2007}, numerical
simulations were used to study the statistics and dynamics of supercontinuum rogue waves
in more detail \cite{Genty-2008}. These simulations showed that the distinct solitons
observed in the picosecond supercontinuum emerged from an initial phase of noise-driven
modulation instability, and the shifts to longer wavelengths of the small number of
``rogue solitons'' in the filtered tail of the DFT histogram arose from inelastic
collisions mediated by the Raman effect \cite{Genty-2008,Mussot-2009}. However, although
certainly leading to long-tailed statistics, these Raman soliton dynamics appeared to be
very specific to supercontinuum
generation, and therefore of primary relevance to the optical domain \cite{Dudley-2014}.

On the other hand, the initial propagation phase of modulation instability had earlier been
proposed as a mechanism for hydrodynamic rogue wave generation
\cite{Trulsen-1997,Onorato-2000,Pelinovsky-2001,Dyachenko-2005,Zakharov-2006}, and
attention in optics therefore quickly focussed on this regime.
In fact, modulation instability and solitons in fibre
optics had been studied since the 1980s, but had been characterized using only
time-averaged measurements of the optical spectrum and/or the temporal intensity
autocorrelation function \cite{Mollenauer-1980,Tai-1986,Taylor-2005,Agrawal-2012}.
These were of course state-of-the-art characterisation techniques for the time, but
the development of real-time techniques such as the DFT, and the possibility of an
analogy with rogue waves, gave new impetus to experimental studies in optics.

There was particular interest in using optics to
study nonlinearly-localized analytic NLSE ``breather'' structures that
emerged from modulation instability, as it had been suggested that their nonlinear
growth and decay were typical characteristics of ocean rogue waves
\cite{Dold-1986,Dysthe-1999,Akhmediev-2009}. Moreover, there was also some evidence from
the shape of time-averaged spectra (specifically the slope of the spectral wings) that
breather structures were present in noise-driven modulation instability
\cite{Dudley-2009}.

The first such studies of nonlinear breathers in the context of rogue waves used coherently-modulated fields injected into optical fibre to excite specific
analytic solutions such as shown in Fig. 1a. Because of the
coherent seeding, the excited breathers were stable, and did not require
real-time characterization. Rather, it was possible to use
averaging techniques such as frequency-resolved optical gating (FROG)
\cite{Trebino-2002} or optical sampling \cite{Andrekson-2007} to record the breather
profiles.  These experiments characterised a range of nonlinear structures including the Akhmediev breather, the Peregrine soliton, and the
Kuznetsov-Ma soliton \cite{Kibler-2010, Hammani-2011, Kibler-2012}, and were significant
in motivating studies in hydrodynamic wave tanks as we describe below
\cite{Chabchoub-2011}.

The comparison of these measurements with analytic predictions was particularly
important in showing that an optical fibre system prior to the
onset of supercontinuum generation could be considered as a close-to-ideal NLSE
environment. Additional experiments coherently exciting higher-order breather solutions of the NLSE further supported this interpretation \cite{Erkintalo-2011,Frisquet-2013,Kibler-2015}.  Related experiments
extended the study of hydrodynamic analogies in optics even further, reporting controlled shock dynamics \cite{Wetzel-2016}, and  dam-breaking phenomena \cite{Xu-2017a,Audo-2018}.

Although impressive in showing the ability of an NLSE system to support nonlinear localization, the direct relevance of optical experiments using coherent initial conditions to noise-driven rogue waves on the ocean remained somewhat unclear.  Certainly numerical studies of noise-driven modulation instability show the emergence of random localized structures as shown in Fig. 1b, and a detailed study of the intensity profiles showed that they in fact cluster around the analytic breather solutions of the NLSE \cite{Toenger-2015}.  The key question, however, was whether these could be measured -- such experiments appeared highly challenging, requiring the measurement of real-time temporal profiles with sub-picosecond resolution, much shorter than the response time of available photodetectors.

To this end, new experimental techniques were developed \cite{Narhi-2016, Suret-2016}, and Fig. 3a shows the principle involved. The idea here is to transpose the action of an optical lens that magnifies an object in space, to temporally-stretch noisy picosecond structures from modulation instability to nanosecond duration replicas
that could be measured (in real time) using high speed photodetectors.  Such a ``time lens'' exploits the mathematical equivalence of paraxial diffraction in space and linear group velocity dispersion in time \cite{Kolner-1989}, and combines two segments of dispersive propagation on either side of a time lens element that introduces a temporal quadratic phase \cite{Salem-2013}. The addition of an additional heterodyne detection stage can also be used to yield intensity and phase information \cite{Tikan-2018}.

Real-time measurements of noise-seeded NLSE propagation in optical fibre are
shown in Fig.~3 b-c.  Figure 3b shows results using a narrowband noisy
continuous wave input, revealing the emergence of random localized structures with
intensity profiles well fitted by analytic NLSE breather solutions \cite{Narhi-2016}.
The measurement of a large data set also allowed the intensity statistics of the modulation instability peaks to be directly characterised, confirming the expected highly-asymmetric probability distribution \cite{Toenger-2015}. These results particularly highlighted the role of breather collisions (or higher-order breathers \cite{Erkintalo-2011,Frisquet-2013}) in generating the largest intensity events that satisfied statistical criteria to be identified as rogue waves (events above the rogue wave intensity $I_{RW}$.)

The experimental results in Fig. 3c (using a heterodyne time lens) show intensity and phase of spontaneous structures observed in the turbulent regime excited by an incoherent input field with near 100\% intensity fluctuations \cite{Tikan-2018}.  The dynamics in this case are described in terms of incoherent background-free higher-order solitons (see Fig. 1c). Long-tailed intensity statistics were also observed in this study, and were important in confirming that spontaneous rogue-wave like statistical behaviour could be observed with large initial fluctuations beyond the scenario of modulation instability.  Earlier results in a similar turbulent regime \cite{Suret-2016} were also notable for showing how the temporal compression seen during incoherent higher-order soliton propagation led to the emergence of the Peregrine soliton, a remarkable result that was later confirmed using coherent initial excitation \cite{Tikan-2017}.  These experiments also motivated follow-up work to tailor the incoherent optical input spectrum to match a scaled version of the JONSWAP spectrum for ocean waves \cite{Koussaifi-2018}, and a quantitative comparison between the scaled statistics obtained in optics and  those from measurements in a one-dimensional wave tank showed good agreement \cite{Onorato-2006}.

\subsection{Linear and Nonlinear Caustics in Two Transverse Spatial Dimensions}

We have so far described optical fibre experiments studying nonlinear focussing dynamics analogous to one-dimensional wave propagation on deep water.  It is important to note, however, that rogue waves on the ocean have also been attributed to other mechanisms -- the linear superposition of random waves propagating in different directions \cite{Longuet-1957}, linear and nonlinear caustic formation \cite{Peregrine-1979,Brown-2001}, and linear directional focussing \cite{Fochesato-2007,Dudley-2013}.

In optics, experiments on linear wave-shaping mechanisms have been performed by studying laser speckle.  Laser speckle is the granular intensity distribution that arises from the spatial interference of a large number of coherent wavefronts with random phases \cite{Goodman-1976}, and is known to exhibit long-tailed statistical properties \cite{Bromberg-2014}.  To study the potential link between speckle and rogue waves, the experiments in Ref.~\cite{Mathis-2015} imprinted a random phase pattern on the transverse profile of a laser beam, and the variation in the beam profile was then characterized during linear propagation (diffraction) in air.  With sufficient propagation, the random initial phase was converted into amplitude fluctuations across the transverse  profile and, using phase retrieval techniques, it was possible to determine the statistics of both the spatial amplitude and intensity. By controlling the strength of the initial random phase, the far-field intensity pattern varied from partially-developed speckle to a broadband caustic network \cite{Nye-1999}.  In the latter case, a long-tailed distribution was observed with a significant fraction of rogue waves. In an intermediate range of initial phase fluctuations, it was possible to synthesize an ``optical sea'' where the spatial amplitude statistics followed a Rayleigh distribution but still showed the presence of a small fraction of events above the rogue wave threshold \cite{Mathis-2015}.  These results are shown in Fig. 3d, and were important in explicitly confirming the possibility to see rogue wave statistics from purely linear propagation in an optical system displaying a clear analogy with ocean wave superposition. It is also worth noting in this context that rogue waves have been shown to arise from purely linear superposition in a one-dimensional environment, for a sufficiently large number of random interacting data bits in an optical communications channel \cite{Vergeles-2011}.

A follow-up experiment studying optical caustics (again using random phase fluctuations as initial conditions) replaced linear propagation in air with nonlinear propagation in a gas cell described by a two-dimensional NLSE model \cite{Safari-2017}.  In this case, experiments showed that the effect of nonlinearity was to enhance the generation of high intensity caustics on the transverse beam profile as shown in Fig. 3e.  The particular significance of these results is that they showed that nonlinearity could enhance the generation of high intensity events whose initial origin was due to a linear propagation effect.   This underlines clearly in the optics domain that both linear and nonlinear focussing can combine in the generation of rogue waves.

\begin{figure}
\begin{center}
\includegraphics[angle=0,origin = c, scale=0.8]{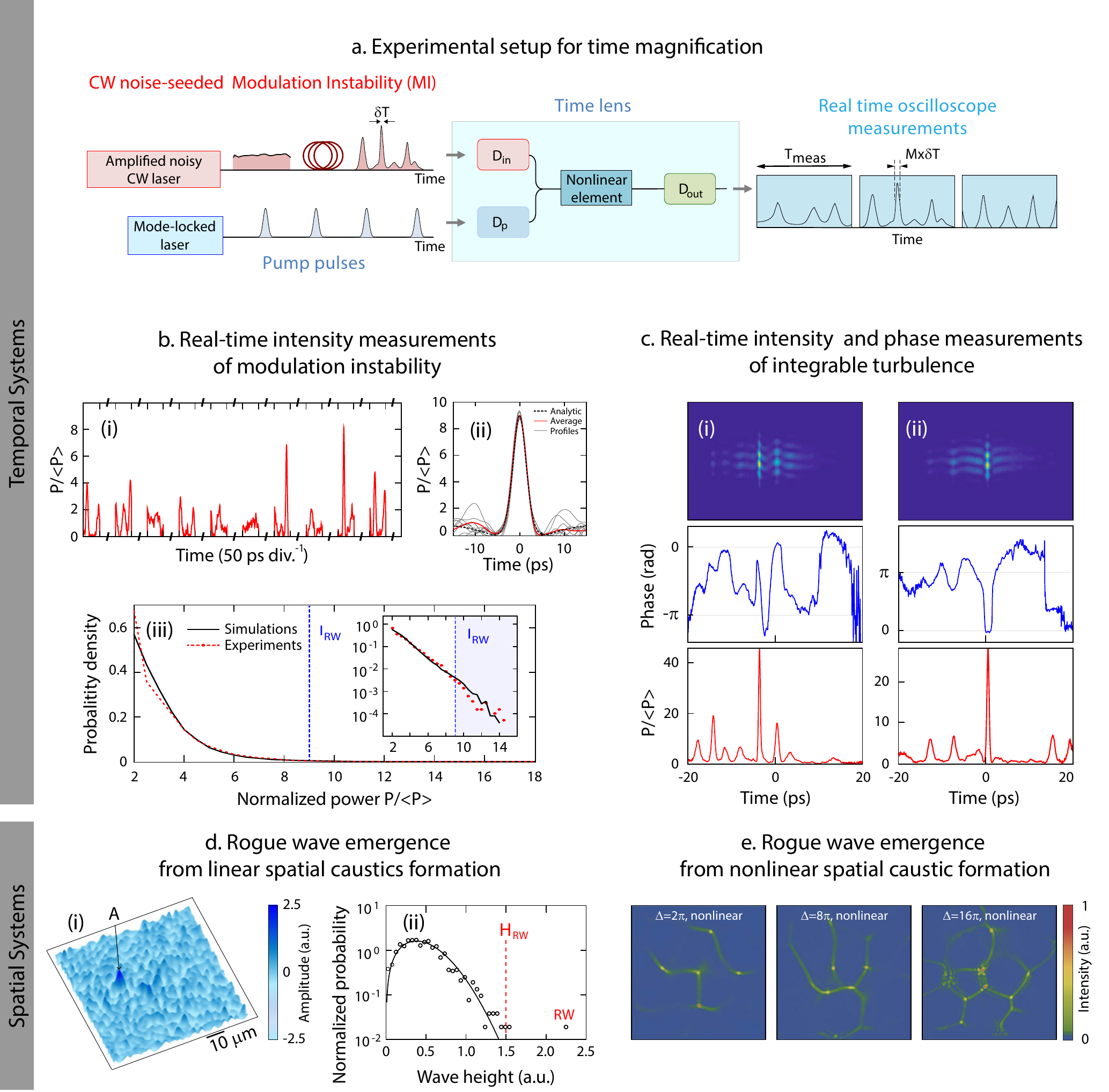}
\caption{Optical rogue wave measurements. \textbf{a}. Schematic for temporal magnification of random modulation instability generated by a noisy CW laser injected in optical fibre. Pulses from a mode-locked laser develop a quadratic temporal phase after dispersive propagation $D_{\rm p}$ before being combined with the dispersed modulation instability signal in a nonlinear element (waveguide or crystal). The nonlinear element transfers the pump quadratic phase to the modulation instability signal such that another dispersive propagation step yields time-magnification by a factor $M = \left|D_{\rm in}/D_{\rm out} \right|$ where $D_{\rm in}$ and $D_{\rm out}$ are the dispersion parameters before and after the nonlinear element. \textbf{b}. Real-time measurements of breather dynamics emerging from noise-seeded modulation instability. (i) Sequential windows of breather intensity profiles normalised with respect to the average output background power $\langle P \rangle$ and rescaled to account for temporal magnification. The measurement window $T_{\mathrm{meas}}$ is determined by the mode-locked pulses (note the broken axis between measurements). (ii) Superposed experimental breather profiles (grey lines) comparing their average (red line) with the analytic Peregrine soliton (black dashed line). (iii) Long tailed normalized probability density of random modulation instability breather peaks, comparing experiments (red) and simulations (black). The inset uses a semi-logarithmic axes. The calculated rogue wave intensity threshold ($I_{\rm RW}$) is shown as a dashed blue line. \textbf{c}. Real-time complex field (intensity and phase) of integrable turbulence dynamics generated by nonlinear fibre propagation of a broadband noisy input field with near 100\% contrast. The measurements use a time-lens setup with the addition of heterodyne detection for phase retrieval. The figure shows: typical raw images captured by the setup (top panels) and retrieved phase (middle panels) and intensity (bottom panels). \textbf{d}. Rogue wave statistics in a purely linear optical system due to caustic focussing of a spatial field with random phase. (i) Measured spatial amplitude of the electric field when the initially applied random phase yields a spatial spectrum intermediate between that of a partially-developed speckle and a caustic network. The trough-to-crest statistics are near-Rayleigh distributed (fit showed as the black solid line) with an extended tail including rogue wave (RW) events exceedingthe significant wave height ($H_{\rm RW}$). The bright peak A labelled in (i) is the highest peak observed in the distribution. \textbf{e}.  Typical caustic networks observed during spatial rogue wave generation in a Rubidium cell via nonlinear amplification of initial small phase fluctuations. Panel \textbf{b} is adapted from Ref. \cite{Narhi-2016}, Springer Nature Limited. Panel \textbf{c} is adapted from Ref. \cite{Tikan-2018}, Springer Nature Limited. Panel \textbf{d} is adapted from Ref. \cite{Mathis-2015}, Springer Nature Limited. Panel \textbf{e} is adapted from Ref. \cite{Safari-2017}, APS.}
\end{center}
\end{figure}

\subsection{Transient Instabilities in Lasers}

In addition to the experiments described above where a close analogy exists with ocean waves, recent work in optics has seen the development of an active area of research where no such analogy has yet been found -- the characterization of ultrafast transient instabilities in modelocked lasers.  In particular, although stable modelocked lasers are well-known to produce highly regular pulse trains, they can also exhibit complex behaviour during their start-up dynamics or when detuned from steady state \cite{Haken-1986}.  Moreover, recent years have seen modelocked laser instabilities described in terms of the wider field of dissipative soliton theory, where the localized soliton concept is generalized from a balance between only dispersion and nonlinearity to also include dissipation in the form of gain and loss \cite{Grelu-2012}.  \jdedit{This has given rise to the description of extreme instabilities in such lasers as dissipative rogue waves \cite{SotoCrespo-2011}.}


Following the  2007 optical rogue wave experiments in fibre, experiments using pulse
energy measurements showed that modelocked lasers could also exhibit long-tailed statistics \cite{Kovalsky-2011,Lecaplain-2012}.
The real time DFT technique was also used to study fibre laser spectral instabilities
\cite{Runge-2013}, yielding measurements of a novel soliton collapse and recovery (or ``explosion'') regime \cite{Runge-2015, Liu-2016}. Typical results are shown in Fig. 4a.  Interestingly, earlier photodiode measurements and experiments using FROG \cite{Dudley-1999} or single-shot spectral characterization \cite{Cundiff-2002} had reported signatures of complex instabilities in modelocked lasers, but the availability of the DFT and time lens techniques significantly renewed interest in this field.

Studies were also performed on the soliton build up
dynamics in a Kerr-lens modelocked Ti:Sapphire laser \cite{Herink-2016}.  Figure~4b shows results from a modified DFT method used for real-time spectral interferometry characterization of soliton
bound states \cite{Herink-2017}. Many other recent experiments have used the DFT technique to study evolving soliton dynamics and multipulse states
\cite{Yu-2017,Liu-2018,Sun-2018,Hamdi-2018,Du-2018,Wei-2018,Wang-2018,Suzuki-2018}, as well as
intensity and coherence measurements in partial and coherent modelocking regimes
\cite{Xu-2016}. A particular example of dissipative soliton formation in a polarization rotation mode-locked fibre laser \cite{Peng-2018} is shown in Fig. 4c. A particular advantage of real-time spectral measurements is that it allows complementary information on temporal soliton separation to be determined through the associated field autocorrelation as shown in both Figs 4b and c.

Combining realtime spectral characterisation using DFT and high speed measurements using
photodiodes allowed the study of a range of multi-scale dynamics in a modelocked fiber
laser \cite{Wei-2017}.  By combining the DFT simultaneously with the time-lens
technique, simultaneous measurement of spectral and temporal profiles was possible with
sub-nm and sub-ps resolution. These experiments yielded a complete picture of
the unstable startup dynamics of dissipative solitons in a modelocked fibre
laser \cite{Ryczkowski-2018}, with typical results showing the growth and collapse of a temporal multi-soliton complex shown in Fig. 4d(i).   Moreover, with simultaneous measurement of spectral
and temporal intensity profiles, the use of phase retrieval techniques allowed
reconstruction of the full field (in amplitude and phase) of the evolving solitons as shown in Fig. 4d(ii).  Finally in this section, we note that in addition to studying modelocked lasers, real-time techniques have also been applied to study
related soliton dynamics in other dissipative systems such as microresonators
\cite{Li-2017,Anderson-2017}.

\begin{figure}
\begin{center}
\includegraphics[angle=0,origin = c, scale=0.8]{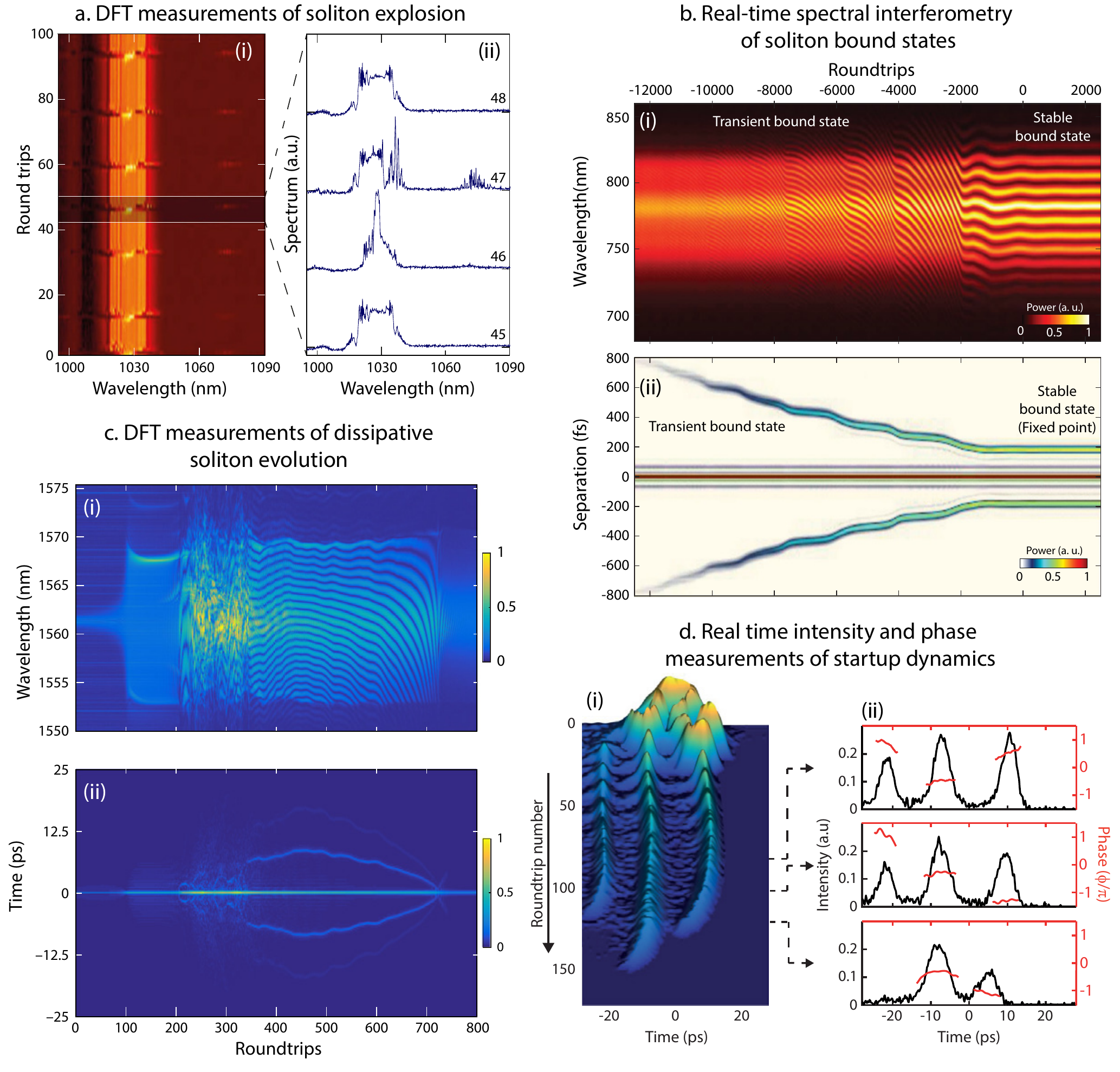}
\caption{\textbf{a}. Dissipative soliton explosions in an Yb-doped mode-locked fiber laser operating in the transition regime between stable mode-locking and noise-like emission. (i) Experimental single-shot spectra of 100 roundtrips showing several explosion events.  (ii) A close-up of the explosion dynamics where the spectrally broad soliton collapses into a narrower spectrum with higher amplitude before recovering to its previous state. \textbf{b}). Soliton bound states in a few-cycle mode-locked laser. (i) Real-time interferogram of 15,000 consecutive cavity roundtrips showing soliton bound state formation with locked phases. (ii) Field autocorrelation evolution over the 15,000 round trips, showing reduction of the soliton temporal separation to form a stable bound state. \textbf{c}. Formation of coherent dissipative soliton structures from unstable noise in nonlinear polarization rotation mode-locked fibre laser. (i) The real-time spectral evolution of the laser output during dissipative soliton build-up measured with DFT. (ii) The field autocorrelation evolution over 800 round trips, tracing the evolution of the temporal separation between the dissipative solitons. \textbf{d}. Dissipative soliton dynamics during start-up phase of a passively mode-locked Er-fiber laser. (i) Temporal evolution over 170 roundtrips captured using a time-lens system showing growth and decay of multiple dissipative soliton structures as the laser passes through a transient unstable regime before stable mode-locking. (ii) Plots showing the full field (intensity in black and phase in red) reconstructed from simultaneous dispersive Fourier transform and time-lens measurements, corresponding to specific roundtrips as indicated.  Panel \textbf{a} is adapted from Ref. \cite{Runge-2015}, OSA. Panel \textbf{b} is adapted from Ref.\cite{Herink-2017}, AAAS. Panel \textbf{c} is adapted from Ref. \cite{Peng-2018}, Springer Nature Limited. Panel \textbf{d} is adapted from Ref. \cite{Ryczkowski-2018}, Springer Nature Limited.}
\end{center}
\end{figure}

\section{Rogue waves in oceanography}

\subsection{Overview}

Although stories of unexpected large ocean waves date back to antiquity, it was really only during the 20th century that any comprehensive scientific study of their properties began \cite{Kharif-2008}. Indeed, several catalogues of records of rogue wave are now providing insights into the long history of attempts to record their occurrence as distinct oceanic events, as well as to understand their properties \cite{Liu-2007,Nikolkina-2011,Nikolkina-2011a,OBrien-2013,OBrien-2018}.  The description of such events as ``freak waves'' was apparently first used in the scientific literature in 1964 \cite{Draper-1966}, although this terminology is to be found in newspaper accounts even earlier \cite{Aberdeen-1951}.  Interestingly, the alternative (now more common) appellation ``rogue wave'' seems to have first appeared in a 1962 novel by C.~S.~ Forester \cite{Forester-1962}, and although a work of fiction, we nonetheless read a remarkably clear physical description of `the ``rogue wave,'' generated by some unusual combination of wind and water'.

Such historical references aside, it is generally accepted that the systematic study of rogue waves began with the milestone experimental measurement on 1 January 1995 of a wave with trough-to-crest height of 25.6~m on the uncrewed Draupner E oil platform in the North Sea \cite{Haver-2004}.  The fact that this Draupner (or New Year's) wave was observed within an extended time series at high sampling frequency yielded great confidence in the measurement fidelity, and stimulated the oceanographic community to investigate the physics and statistics of ocean rogue waves on a quantitative level.

This renewed interest in rogue waves was able to build upon an impressive literature of existing theoretical and experimental work.  Although the ocean environment involves multiple processes such as currents, dissipation, wind forcing, and wave breaking, early studies had laid the groundwork for understanding large amplitude wave shaping dynamics in terms of both constituent linear and nonlinear mechanisms.  For example, the statistics of random waves had been linked to the spectral properties of the wave field \cite{Longuet-1957}, linear dispersive focussing had been suggested to generate high amplitude waves \cite{Longuet-1974}, and nonlinear focussing due to modulation instability in the NLSE had been the subject of a large number of theoretical studies  \cite{Lighthill-1965,Whitham-1965,Bespalov-1966,Benjamin-1967a,Zakharov-1968,Peregrine-1983}.

It is reasonable to say that it has been nonlinear focussing from modulation instability that has attracted most recent attention within the oceanography community as a potential mechanism underlying rogue wave formation.  In particular, because its theoretical derivation arises from the integrable NLSE, it is amenable to analytic study which is attractive in trying to develop a physical picture of the underlying dynamics.  And as we have discussed above, it was the analogy with nonlinear propagation in fibre which stimulated renewed interest in rogue waves in a wider physical context.  That said, the relative contribution of linear and nonlinear effects to drive rogue wave dynamics in the ocean remains a subject of much study \cite{Adcock-2014, Onorato-2016}, and in our discussion below of wave tank experiments and analysis of real-world ocean wave data, we take particular care to highlight when results have drawn conclusions regarding the roles played by linear and/or nonlinear effects.  This is especially the case for recent analyses of real world ocean data that have been able to reproduce rogue wave characteristics through spatio-temporal effects without the need to include  nonlinear focusing from modulation instability.

\subsection{Rogue waves in the natural environment}

Because rogue waves are rare events with sudden growth and decay dynamics, their measurement is extremely challenging, requiring both long time series (over months or even years) to capture their rarity, and a sampling frequency (between 2-10~Hz) capable of recording the main features of wave shapes.  The rogue wave measurements that have been assessed as most reliable are arguably those based on the reflection of an optical or acoustic signal at the boundary between sea and air \cite{Kharif-2008}.  The 1995 Draupner wave for example was recorded by a downward looking laser sensor using a sampling frequency of 2.13~Hz during 20 minutes of every hour. A similar technique was used in recording a second landmark measurement of what is known as the Andrea wave,  measured on 9 November 2007 in the North Sea using a  sampling frequency of 5~Hz during 20 minute intervals by 4 independent lasers mounted on the Ekofisk platform \cite{Magnusson-2013}.  Related measurement techniques include using mounted microwave radar \cite{Christou-2014} and acoustic Doppler current profilers \cite{Flanagan-2016}.

Other methods for measuring rogue waves have been reported, but there has been significant debate regarding their fidelity.  For example, measurements relying on accelerometers placed on wave buoys can exhibit distortion due to the buoy's intrinsic moment of inertia \cite{Kharif-2008}, although a number of rogue wave events have been extracted from buoy measurements using appropriate quality control \cite{Pinho-2004,CasasPrat-2010,Baschek-2011,Cattrell-2018}.  Moreover, there is growing appreciation that the majority of measurements to date are limited in the information they can yield about rogue wave properties because they record time series at only the single point where the sensor has been located.

To this end, improved approaches to satellite remote sensing continue to be explored \cite{Ardhuin-2018} and techniques such as stereo video have been shown to be especially promising as they allow capture of the space-time evolution of the sea surface over an extended region \cite{Gallego-2011,Benetazzo-2017}.  There has also been significant interest in remote sensing using satellite-based synthetic aperture radar (SAR) \cite{Lehner-2004}, but such imagery has been shown to be associated with errors due to velocity bunching and azimuthal image smear \cite{Janssen-2006}.   Nonetheless, there is emerging consensus that recording the full spatio-temporal evolution of waves on the ocean's surface is essential to fully characterize the statistics of large waves in complex sea states such as hurricanes and storms.  This is because the maximum height of a group of waves propagating in a complex manner over a large area during a given time interval will likely be greater than the wave height observed at only one fixed point in space (as measured by e.g. a moored buoy). The study of such ``space time extremes'' is an important current area of research \cite{Benetazzo-2017, Fedele-2017}.

Despite the challenges of measurement, appropriate quality control has been applied to many extended wave record time series, and potentially 1000's of candidate rogue waves have been identified  \cite{Kharif-2008}.  A summary of various results obtained from fixed platforms as well as an analysis of a large dataset of wave buoy data has recently appeared in Ref.~\cite{Cattrell-2018} and a particular conclusion of this summary was that mechanisms for rogue wave formation vary from place to place on the ocean.

The characteristics of several large ocean wave events have also been analyzed in detail using an approach known as hindcasting.  Such studies use archived meteorological and wave data at the location of the event under study to determine initial conditions for a forward-propagating wave model \cite{Cardone-1976}.  The aim is to use the model to simulate the wave field characteristics at later times and compare quantitatively with a measured wave record.  By varying the initial conditions and parameters of the model, it is possible to draw conclusions as to which particular processes may be responsible for the emergence of the observed rogue waves.  Recent work of this type includes analysis of the maritime accidents of the cruise ship Louis Majesty \cite{Cavaleri-2012}, the tanker Prestige \cite{Trulsen-2015}, the merchant vessel El Faro \cite{Fedele-2017}, as well as modelling of the Draupner and Andrea waves \cite{Fedele-2016,Cavaleri-2017}, and rogue waves generated during Typhoon Lupit in 2009 \cite{Fujimoto-2018}.  The particular case of the Andrea wave and its modelled profile using the hindcasting approach is shown in Fig.~5a.

Significantly, these studies of individual examples of real-world ocean waves have highlighted the difficulty in identifying a single primary cause for the wave enhancement, because different approaches to analysing the same event can yield different conclusions. For example, motivated in part by the presence of crossing sea states at the time of the Draupner wave, the analysis in Ref. \cite{Onorato-2006a} initially suggested that third-order nonlinear focussing (modulation instability) in crossing seas may be important in rogue wave formation.  But further analysis and direct numerical modelling of the Draupner wave with and without the cubic nonlinear term in simulations showed that such third-order nonlinear focussing was negligible \cite{Fedele-2016}.  In the latter case, the wave shaping was found to be dominated by directional superposition, with a minor enhancement from the second-order bound (Stokes) nonlinear contribution (a similar conclusion applied to the Andrea wave shown in Fig.~5a.)  In this regard, we also note that a negligible role for third-order nonlinear focussing was identified from the hindcasting analysis in Ref. \cite{Trulsen-2015} considering the wave conditions associated with the Prestige accident.

Similar results were obtained by \cite{Gemmrich-2017} in a study of 2~million wave groups where they identified 300 rogue waves. Although nonlinear modulation instability is discussed as a possible effect that can increase wave envelope steepness, the conclusion (based on the symmetric shape of the wave groups) was that random superposition of the Stokes waves was sufficient to explain the observations of individual rogue waves. Results from several other groups have independently supported the interpretation that downplays the role of modulation instability and instead highlights the role of linear interference and/or localized dispersive focusing \cite{Christou-2014, Benetazzo-2017}.  On the other hand, the hindcasting analysis of rogue waves observed during Typhoon Lupit suggested that focussing nonlinearity played a role in some instances of rogue wave formation \cite{Fujimoto-2018}.

When these various results are taken together, it appears that the complexity of the ocean coupled with the relatively limited observational data suggest that a definitive conclusion concerning the underlying physics of ocean rogue wave formation may be premature.  And as we discuss more below in the context of laboratory experiments, both linear and nonlinear processes remain under active study as potential rogue wave driving mechanisms. In this regard, there is naturally significant interest in searching for convenient statistical signatures (in both simulation and real-world wave data sets) that may allow the role played by either linear or nonlinear processes to be distinguished.  Of particular promise here is the fourth order moment (kurtosis) which has been shown to provide a robust measure of the strength of the tails of a skewed wave height distribution \cite{Janssen-2003,Annenkov-2009,Fedele-2015}.

\begin{figure}
\begin{center}
\includegraphics[angle=0,origin = c, scale=0.55]{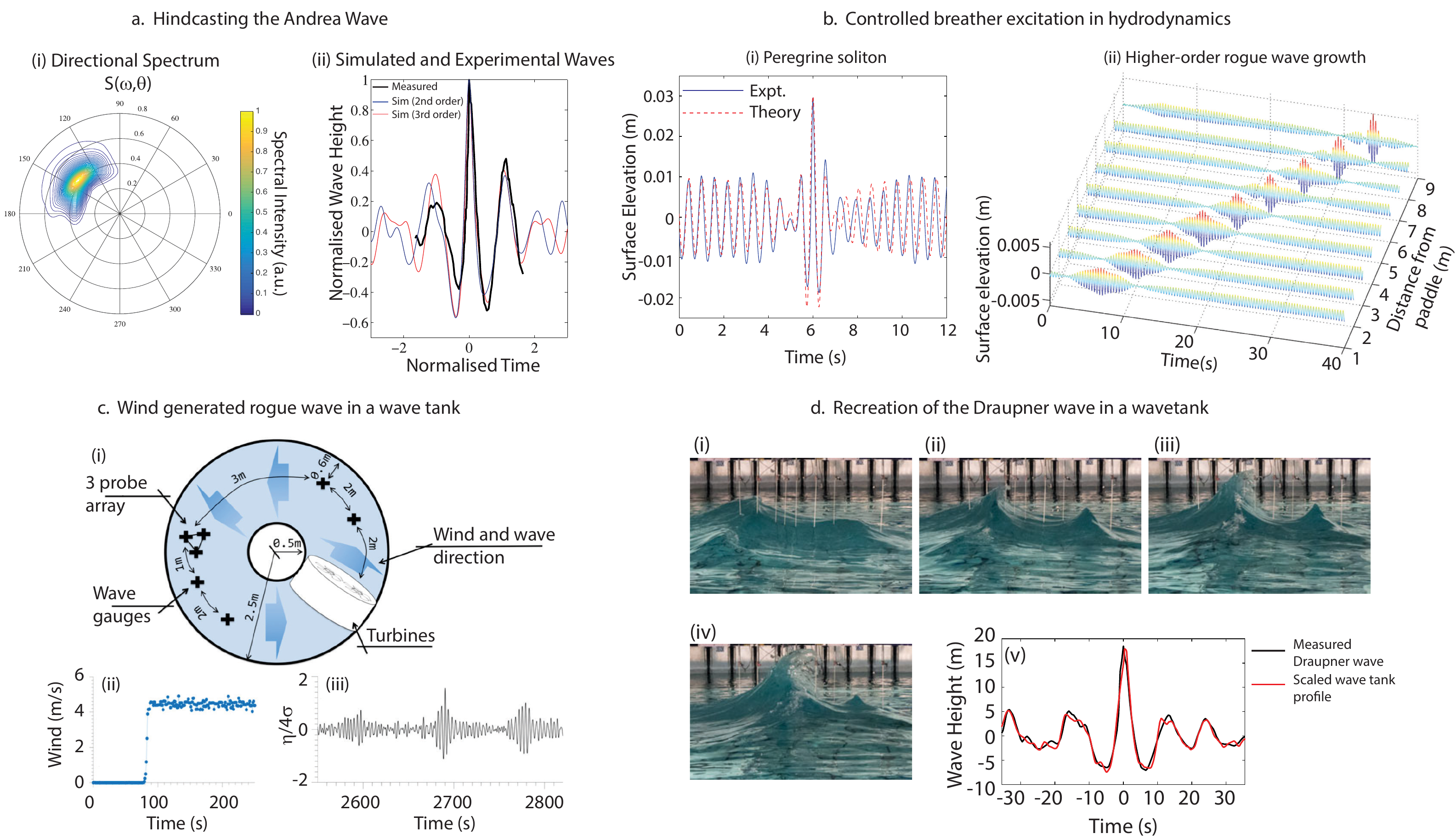}
\caption{Ocean and wave tank rogue wave measurements. \textbf{a}. Hindcasting results for the Andrea wave. (i) The directional spectrum $S(\omega,\theta)$ used in numerical simulations.  (ii) The measured Andrea wave (black) compared with simulations including up to second-order (blue) and third-order contributions (red).  The minor differences between the simulations point to a negligible role from third-order nonlinear focussing.  \textbf{b}.  Controlled breather generation in wave tanks using initial conditions in amplitude and phase determined from analytic breather solutions to the NLSE. (i) Experimental (blue) Peregrine soliton compared with the analytic prediction (red).  (ii) Measured evolution in a wave tank of a higher-order breather solution to the NLSE.  \textbf{c}.  Wind-generated waves in an annular wave tank.  (i) Schematic of the elements of the experimental setup. (ii) Example of wind speed profile used for wave generation.  (iii) Example of measured water surface elevation ($\eta/4\sigma$ i.e. normalised to four times the standard deviation of the 10 minute record).  The wave record shown includes a rogue wave of height 2.7 time higher than the associated significant wave height.  \textbf{d}.  Recreation of the Draupner wave in a circular wave tank for crossing seas with a crossing angle of $120^\circ$.  (i)-(iv) show images of the free surface taken at intervals of 100~ms.  (v) Comparison of the scaled wave tank reproduction (red) with the measurements at the Draupner platform (black).  Panel \textbf{a} is adapted from Ref. \cite{Fedele-2016}, Springer Nature Limited. Panel \textbf{b} is adapted from (i) Ref. \cite{Chabchoub-2011}, APS and (ii)\cite{Chabchoub-2012}, APS.  Panel \textbf{c} is adapted from Ref. \cite{Toffoli-2017}, APS. Panel \textbf{d} is adapted from Ref. \cite{McAllister-2019}, APS.}
\end{center}
\end{figure}

\subsection{Wave tank experiments}

Although the aim of any study of rogue waves is to gain insight into their properties on the ocean's surface, reproducing extreme wave propagation in laboratory wave tanks has been very important in allowing comparisons to be made with theory and modelling under controlled conditions.  Of course, for wave tank experiments to have relevance to ocean wave dynamics, the experimental conditions should mimic as far as possible the natural processes.  To this end, state-of-the art wavemakers can be conveniently programmed to generate a wide range of  initial conditions, and with artifacts due to effects such as reflection and viscosity well-understood, it is possible to use wave tanks to study a wide range of propagation scenarios in deep or shallow water of constant or varying depth.  Although many experiments on one dimensional propagation can be performed in narrow channels or flumes, the use of wave-making panels on the perimeter of larger wave basins, or the use of circular wave tanks is now allowing convenient study of multi-dimensional effects \cite{Gyongy-2014}. Another geometry recently developed for wave tank experiments is the annular flume that can allow the study of some wave generation phenomena through circular propagation under essentially unlimited distance (fetch) \cite{Toffoli-2017}.

Interestingly, the first use of wave tanks to study modulation instability actually arose from an experiment that went wrong \cite{Hunt-2006}.  Aiming to test the stability of small amplitude water waves experimentally, Benjamin and Feir unexpectedly observed the exponentially-growing amplitude modulation of the wave train \cite{Benjamin-1967b}.  Following this pioneering work, other wave tank experiments investigated related deep-water (focussing nonlinearity) propagation effects such as Fermi-Pasta-Ulam-Tsingou recurrence, the emergence of strongly-modulated wavepackets, and the formation of isolated NLSE solitons \cite{Yuen-1975,Lake-1977,Yuen-1982}.  Later experiments examined the effect of wave breaking with particular emphasis on how breaking influences the long-time evolution of an evolving wave train \cite{Rapp-1990,Tulin-1999}. Following the first observation of the Peregrine soliton in optics, there was renewed interest in studying analytic breather solutions to the NLSE in wave tanks with suitably tailored initial conditions.  This led to a number of experiments reporting similar Peregrine soliton dynamics \cite{Chabchoub-2011}, as well as higher-order breathers \cite{Chabchoub-2012}.  The wave profiles measured in these experiments are shown in Fig. 5b (i) and (ii) respectively. An additional study even generated a Peregrine soliton and examined its interaction with a model of a chemical tanker in a seakeeping test \cite{Onorato-2013a}.

A very important experiment looked specifically at how rogue waves could be triggered randomly, by driving a wavemaker with scaled initial conditions replicating the experimental JONSWAP spectrum of waves on the North sea \cite{Onorato-2006}. Experimental results were consistent with the emergence of localised NLSE breathers from modulation instability.  These results supported earlier numerical studies that had predicted that modulation instability would increase the probability of rogue waves in random oceanic sea states \cite{Onorato-2001}, although the authors noted the limitation of their work in that it considered only one-dimensional dynamics.  Significantly, these wave tank experiments have been confirmed using larger statistical data sets in recent experiments in optics where the optical initial conditions were tailored and scaled to match the JONSWAP spectrum parameters used in the wave tank experiments \cite{Koussaifi-2018}.  In another related experiment, a possible rogue wave triggering effect was demonstrated in a wave tank by adding a  coherent Peregrine soliton state to a random JONSWAP wave spectrum as initial conditions to a wavemaker \cite{Chabchoub-2016}.

Other recent work has aimed to include processes found in the natural environment in wave tank experiments.  For example, studies have shown that modulation instability and Peregrine soliton evolution can persist or even be enhanced in the presence of opposing currents \cite{Toffoli-2013a,Liao-2018}.  In other experiments, the effect of wind on rogue wave dynamics has been studied in specially adapted wind-sea wave tanks \cite{Kharif-2007}. Particular experiments in a one-dimensional wave tank have shown that wind can induce frequency downshifting in the spectrum of initial breather solutions excited by a wavemaker \cite{Waseda-1999,Eeltink-2017}, while other experiments in an annular wave tank did not use any initial mechanical wave generation, but allowed waves to develop spontaneously from the action of the wind on the water's surface \cite{Toffoli-2017}. These results are shown in Fig.~5c and are significant in showing that wind forcing could indeed generate conditions for the observation of non-Gaussian long-tailed wave height statistics.

In addition to studies that have explicitly set out to examine the role of modulation instability and related nonlinear focussing on rogue wave dynamics, experiments have also shown that linear dispersive focussing can also generate large amplitude waves.  Indeed, the generation of multi-frequency initial conditions with phases adjusted to lead to a focussed wave group at a prescribed distance in the wave tank has been used in a number of experiments studying wave-breaking effects from large amplitude waves \cite{Greenhow-1982,Dommermuth-1988,Baldock-1996,Alberello-2018}. Of interest in this context is a one-dimensional wave tank experiment that was able to use deterministic linear superposition of component waves to reproduce a scale model of the single-position wave train of the Draupner wave \cite{Clauss-2011}. Some progress has also been made in wave tanks to understand the role of linear and nonlinear focussing in directional seas \cite{Onorato-2009}, including a recent experiment recreating the Draupner wave from  multi-directional superposition \cite{McAllister-2019} as shown in Fig. 5d.

\section{Summary and Outlook}

It should be clear from the above discussions that the last decade has seen many examples of important and mutually-beneficial studies of rogue waves in optics and hydrodynamics.  A particular focus of recent studies has concerned the relative contributions of linear and nonlinear effects in driving rogue wave emergence.  However, although one-dimensional modulation instability processes have been extensively studied in both optics and in wave tank experiments, it is not possible to conclude that such nonlinear focussing is the dominant mechanism underlying extreme wave dynamics on the ocean.  In any case, our view is that it is not useful to focus on any one single cause of all rogue waves. Rather, we believe that an objective interpretation of the current literature leads to a conclusion that rogue waves on the ocean most likely arise from a number of linear and nonlinear processes that contribute separately or in combination depending on the particular ocean conditions at play.

In this regard, we anticipate that future progress in unravelling the complexity of ocean waves will require targeted studies using wave tanks and improved {\em in situ} measurements of ocean waves in their natural environment.  For example, wave tank experiments are proving important in ongoing studies of processes such as dissipation, wavebreaking and air-sea interaction, and there is no doubt that attempts to understand ocean wave dynamics using wave tank experiments will remain a major area of research for decades to come.  And in addition to studies of rogue waves, there are also many studies required to improve understanding of how nonlinear focussing may contribute to other phenomena such as wave run up \cite{Carbone-2013}.  It is also possible in this regard that advances in developing new analogous spatial propagation systems in optics may lead to further areas of cross-fertilization where insights between the two disciplines can continue to be shared.

As well as analogous experiments, there are also new potential areas of overlap between optics and hydrodynamics from the perspective of data analysis. For example, complementing efforts that are developing approaches to deterministic prediction \cite{Osborne-2010,Cousins-2016}, an important emerging area of research in both optics and oceanography is the application of techniques from machine learning to detect patterns and build models based on analysis of large data sets \cite{Jordan-2015,LeCun-2015}. In particular, for complex physical systems where there is no obvious model linking input and output, a machine learning algorithm can be trained using measured or simulated input and output data to determine an effective input-output model that can subsequently be used for predictive purposes \cite{James-2013}.  In optics, such techniques have attracted much attention in areas such as telecommunications and laser stabilisation \cite{Woodward-2016,Zibar-2017,Baumeister-2018}, and very recently have been applied to the analysis of intensity peaks in modulation instability \cite{Narhi-2018}.  There has been similar broad interest in oceanography, with recent applications including the development of advanced statistical analysis of irregular waves \cite{Mohamad-2018},  prediction of tidal currents \cite{Sarkar-2018}, and wave forecasts \cite{ODonncha-2018,James-2018}.

We also note that recent experiments in optics that yielding access to the full electric field of propagating light pulses have allowed explicit calculation of
the corresponding complex nonlinear eigenvalue spectrum \cite{Ryczkowski-2018}.  Although well-known in
mathematical analysis of nonlinear propagation and in studies of ocean rogue waves \cite{Osborne-2010}, the ability to calculate such a nonlinear spectrum from experimental data in optics is a significant advance which has already had applications in fundamental studies of optical turbulence \cite{Randoux-2016}.  Such measurements in optics also fall within a developing field related to applications of nonlinear Fourier transforms as a solution to overcome bandwidth limitations in optical telecommunications \cite{Turitsyn-2017}.

Since the first analogy between optical and ocean rogue waves was proposed in 2007, there have been remarkable developments by groups worldwide, and an active field of interdisciplinary rogue wave physics has now been established.  The open questions that remain and the evident areas of common interest lead us to expect increased fruitful interactions between these two disciplines.

\newpage

\noindent\textbf{Rogue Wave Glossary}

\vskip 1cm

\noindent
{\sc Akhmediev Breather.}  A soliton on finite background solution to the NLSE that describes a single cycle of growth and decay along the propagation direction $\xi$ with periodic behaviour along the time axis $\tau$.
\vskip 2mm

\noindent
{\sc Carrier oscillations.} Individual cycles of a propagating wave underneath a group or pulse envelope.
\vskip 2mm

\noindent
{\sc Coherence.} Phase stability of carrier oscillations of a single frequency wave, or the stability of the phase difference between the carrier oscillations of two waves.
\vskip 2mm

\noindent
{\sc Continuous Wave or CW.} Also known as ``plane wave'', this is a wave of constant amplitude or intensity.
\vskip 2mm

\noindent
{\sc Crossing Seas.} A sea state with two independent wave systems travelling at oblique angles.
\vskip 2mm

\noindent
{\sc Directional wave energy spectrum.} The distribution of wave energy in frequency and direction, often used to provide initial conditions for multi-dimensional linear and nonlinear wave modelling.
\vskip 2mm

\noindent
{\sc Dispersive Fourier Transform.} Also known as Time Stretch and used for real-time spectroscopy, this technique temporally stretches an ultrashort pulse through linear dispersion such that its temporal intensity assumes the form of its spectrum.
\vskip 2mm

\noindent
{\sc Dissipative Soliton.} Stable localized structure that is localized as a result of balance between nonlinearity, dispersion and energy exchange (gain or loss) with an environment.
\vskip 2mm

\noindent
{\sc Envelope or Group.} A slowly-varying function that modulates the amplitude of optical or water carrier waves. 
\vskip 2mm

\noindent
{\sc Hindcasting.}  Also known as backtesting, an approach used to test a mathematical model and predict wave elevation properties based on archival inputs such as directional wave energy spectra at an earlier time.
\vskip 2mm

\noindent
{\sc Kuznetsov-Ma soliton or breather.}  A soliton on finite background solution to the NLSE describing periodic oscillation along the propagation direction $\xi$ with localization along $\tau$.
\vskip 2mm

\noindent
{\sc Long-crested and short-crested waves.}  The ``crest'' of a wave is equivalent to its transverse extent, with ocean waves classified as long-crested or short-crested respectively depending on whether they predominantly propagate in one direction or consist of a superposition of waves propagating in different directions.
\vskip 2mm

\noindent
{\sc Long-tailed distribution.} A characteristic of statistical distributions where the tails decrease very slowly and contain a sub-population of extreme events.
\vskip 2mm

\noindent
{\sc Modulation Instability.} Exponential growth of a weak perturbation on a continuous wave excitation in any nonlinear system. 
\vskip 2mm

\noindent
{\sc Peregrine Soliton.} A limiting case of the Akhmediev breather and Kuznetsov-Ma soliton solutions that is doubly-localised along the propagation direction $\xi$ and $\tau$.
\vskip 2mm

\noindent
{\sc Rogue Wave.} A large amplitude wave satisfying the common definition that its height from trough to crest exceeds twice the Significant Wave Height. In optics, the definition is the same, but expressed in terms of optical intensity.
\vskip 2mm

\noindent
{\sc Significant Wave Height.}  Mean wave height from trough to crest of the upper third of all events in a recorded time series of surface elevation.  In optics the equivalent quantity is Significant Intensity, the mean intensity (from zero) of the upper third of all events in a recorded intensity time series.
\vskip 2mm

\noindent
{\sc Soliton.} A coherent structure in a nonlinear dispersive system that displays either stationary or recurrent behaviour with propagation, including stationary background-free sech-solitons and solitons on finite background that are also known as breathers.
\vskip 2mm

\noindent
{\sc Space-Time Extreme.} The maximum wave surface height observed over a given area during a time interval, and not just at a given point.
\vskip 2mm

\noindent
{\sc Steepness.} For water waves only, the steepness is given by $ka = 2\pi a/\lambda$ where $k$ is the wavenumber, $\lambda$ the wavelength, and $a$ the amplitude.
\vskip 2mm

\noindent
{\sc Stokes wave.}  The surface elevation of water waves including ``bound'' harmonic components which to second order is written $\eta(t,z) = a \cos(\theta)+ 1/2 (ka)^2 \cos(2\theta)$ where $\theta = kz-\omega t$ and $k$ is the wavenumber given here for deep water.   
\vskip 2mm

\newpage

\section*{Funding Information}

JMD acknowledges support from the French Investissements d'Avenir programme, project ISITE-BFC (contract ANR-15-IDEX-0003). GG acknowledges support from the Academy of Finland (grants 298463 and 318082).  AM acknowledges support from the Fonds Europ\'{e}en de D\'{e}veloppement Economique R\'{e}gional (project HEAFISY), the Labex CEMPI (ANR-11-LABX-0007) and Equipex FLUX (ANR-11-EQPX-0017) and the French Investissements d'Avenir programme.  FD acknowledges support from Science Foundation Ireland (SFI) under the research project `Understanding Extreme Nearshore Wave Events through Studies of Coastal Boulder Transport' (14/US/E3111).  Earlier but critical financial support to JMD and FD was provided by the European Research Council (ERC-2011-AdG 290562-MULTIWAVE).

\section*{Acknowledgements}

Our understanding of the physics and applications of supercontinuum generation has benefited from collaboration and discussion with numerous colleagues and friends whom we thank, and we hope that they will excuse us for not being able to acknowledge them individually.  We also thank Dr C. Billet for invaluable assistance in figure preparation.

\section*{Competing Interests}

The authors declare no competing interests.

\newpage


\end{document}